\documentclass[%
    reprint,
    aps,
    prl,
    superscriptaddress,
    showpacs,
    floatfix,
    bibnotes,
    amsmath,
    amssymb,
    amsfonts,
    amsthm,
]{revtex4-2}

\usepackage[utf8]{inputenc}     
\usepackage[T1]{fontenc}        
\usepackage{color}              
\usepackage[svgnames]{xcolor}   
\usepackage[inkscapelatex=true]{svg}
\usepackage{dcolumn}            
\usepackage{bm}                 
\usepackage{enumerate}          
\usepackage{enumitem}	        
\usepackage{csquotes}           
\usepackage{empheq}             
\usepackage{dsfont}             
\usepackage{siunitx}            
\usepackage{physics}            
\usepackage{hyperref}           
\usepackage[caption=false]{subfig}
\usepackage{graphicx}           
\usepackage{cleveref}

\usepackage[mode=buildnew]{standalone}

\usepackage{todonotes}

\graphicspath{{figs/}}

\hypersetup{
	colorlinks  = true,
	citecolor   = Orange!80!black,
	linkcolor   = RoyalBlue!60!black, 
	urlcolor    = RoyalBlue!80!black
}
\usepackage{tikz}
\usepackage{xcolor}
\usetikzlibrary{positioning, decorations.pathmorphing, decorations.markings}
\tikzstyle{site} = [circle, shading=ball, ball color=black!40, minimum size=4mm, inner sep=0pt]

\newcommand{\ham}{{H}}

\newcommand{\bdls}[2]{\ensuremath{b^\dagger_{#1 #2}}}
\newcommand{\bls}[2]{\ensuremath{b_{#1 #2}}}
\newcommand{\bleft}[1]{\ensuremath{b_{#1 \leftarrow}}}
\newcommand{\bright}[1]{\ensuremath{b_{#1 \rightarrow}}}
\newcommand{\bdleft}[1]{\ensuremath{b^\dagger_{#1 \leftarrow}}}
\newcommand{\bdright}[1]{\ensuremath{b^\dagger_{#1 \rightarrow}}}
\newcommand{\nls}[2]{\ensuremath{n_{#1 #2}}}

\newcommand{\systemsize}{\ensuremath{L}}
\newcommand{\numbertrajectories}{\ensuremath{N_\mathrm{traj}}}

\makeatletter
\newcommand{\lrarrow}{\mathrel{\vcenter{\hbox{\ooalign{\raisebox{0.5ex}{$\m@th\scriptstyle\rightarrow$}\cr\raisebox{-0.5ex}{$\m@th\scriptstyle\leftarrow$}}}}}}
\makeatother

\begin{document}

\title{Quantum motility-induced phase separation}

\author{Laurin \surname{Brunner}}
\email{laurin.brunner@uni-a.de}
\affiliation{Theoretical Physics III, Center for Electronic Correlations and Magnetism, Institute of Physics, University of Augsburg, D-86135 Augsburg, Germany}

\author{Ricard \surname{Alert}}
\affiliation{Departament de F\'{i}sica de la Mat\`{e}ria Condensada, Facultat de F\'{i}sica, Universitat de Barcelona, Barcelona, Spain}
\affiliation{Universitat de Barcelona Institute of Complex Systems (UBICS), Barcelona, Spain}
\affiliation{Instituci\'{o} Catalana de Recerca i Estudis Avan\c{c}ats (ICREA), Barcelona, Spain}
\affiliation{Max Planck Institute for the Physics of Complex Systems, Dresden, Germany}
\affiliation{Center for Systems Biology Dresden, Dresden, Germany}
\affiliation{Cluster of Excellence Physics of Life, TU Dresden, Dresden, Germany}

\author{Reyhaneh \surname{Khasseh}}
\affiliation{Theoretical Physics III, Center for Electronic Correlations and Magnetism, Institute of Physics, University of Augsburg, D-86135 Augsburg, Germany}

\author{Markus \surname{Heyl}}
\email{markus.heyl@uni-a.de}
\affiliation{Theoretical Physics III, Center for Electronic Correlations and Magnetism, Institute of Physics, University of Augsburg, D-86135 Augsburg, Germany}

\date{\today}

\begin{abstract}
Motility-induced phase separation (MIPS) describes a central clustering phenomenon in active matter systems where particles spontaneously separate into dense and dilute phases even in the absence of interparticle attractive forces.
Recent theoretical and experimental efforts have taken the first steps to extend active matter concepts to the quantum level.
However, whether a genuine quantum analog of MIPS exists and how quantum coherence would compete or cooperate with the dissipative self-propulsion has remained open so far.
Here, we provide evidence that quantum MIPS can occur in a model of active hard-core bosons in one dimension.
Our model yields superlinear number fluctuations characteristic of MIPS, leading to microphase separation with large but finite cluster size.
Adding nearest-neighbor repulsive interactions, we find numerical evidence for restoring genuine phase separation with a divergent correlation length. 
Crucially, the clustered steady states maintain a long-distance quantum coherence, revealing a genuinely quantum feature with no classical counterpart.
These results provide a foundation for exploring quantum MIPS and suggest that the coherence-activity interplay can generate new types of nonequilibrium quantum states.
\end{abstract}

\maketitle

\paragraph{Introduction.}
\label{sec:introduction}

Active matter systems, composed of self-propelled particles that consume energy to drive their motion, have turned into a paradigmatic class of physical many-body models for nonequilibrium collective phases and pattern formation~\cite{Ramaswamy2010, Marchetti2013, Bowick2022, Vrugt2025, Paxton2004, Howse2007, Narayan2007, Deseigne2010, Buttinoni2013, Das2024}. Among the key discoveries is motility-induced phase separation (MIPS)~\cite{Tailleur2008, Fily2012, Redner2013, Buttinoni2013, Cates2015, Geyer2019, Partridge2019, VanDerLinden2019, Shi2020, Zhang2021}, where purely self-propelled motion drives spontaneous separation into dense and dilute phases, even without attractive interactions.
The physics of MIPS can emerge through a generic feedback mechanism: particles slow down at high density, leading to accumulation and further slowing down, ultimately causing phase separation~\cite{Tailleur2008, CatesTailleur2013, Cates2015}.

Very recent theoretical~\cite{VrugtSchrodinger2023, Antonov2025, Antonov2025b} and experimental~\cite{Burgardt2026} efforts have taken the first steps to lift the concept of active matter to the quantum world. At the single-particle level, the landscape remains explorative on how to render individual quantum particles active via engineered dissipation~\cite{Diehl2008, Verstraete2009, Gipouloux2026, Antonov2025b, Antonov2025}, non-unitary quantum walks~\cite{Yamagishi2023}, or microscopic heat-to-motion conversion~\cite{Penner2025}. Recent studies in the many-body regime have already started to identify collective phenomena such as activity-induced phase transitions, ferromagnetism, and synchronization in open quantum settings~\cite{Adachi2022, Kazuaki2024, Nadolny2025}. Most notably, a recent work has demonstrated active quantum flocks~\cite{khasseh2024}—the first compelling evidence that collective motion analogous to classical flocking~\cite{Vicsek1995, Toner1995, Toner2005} can emerge in quantum systems. However, flocks represent only one facet of active matter phenomenology. The other cornerstone—motility-induced phase separation—has so far remained unexplored in the quantum regime.

Here, we show compelling numerical evidence for quantum motility-induced phase separation by introducing a one-dimensional Lindblad model of active hard-core bosons, which combines coherent quantum tunneling with dissipative jump processes analogous to classical run-and-tumble dynamics~\cite{Soto2014, Slowman2016}. Leveraging an exact factorization of the dynamics, we perform numerically exact Monte Carlo wave-function simulations up to system sizes of $L=1000$ lattice sites.
We find that our minimal model exhibits giant number fluctuations characteristic of MIPS, yet we observe only microphase separation with a large but finite cluster size~\cite{Leibler1980, Tjhung2018, Caporusso2020, Adachi2022}.
To stabilize genuine quantum MIPS, we introduce additional nearest-neighbor repulsion, which enhances the high-density slow-down and thereby the self-trapping mechanism.
In the regime of nearest-neighbor interactions, the characteristic length scale for clusters grows strongly with system size, consistent with genuine phase separation. Most significantly, both models exhibit pronounced quantum coherence in their clustered steady states—a feature absent in classical MIPS.
This persistence of coherence establishes quantum MIPS as a distinct nonequilibrium phase, demonstrating that activity and quantum coherence can cooperate to shape new classes of quantum many-body states.


\begin{figure}[t]
    \centering
    \includegraphics[width=\linewidth]{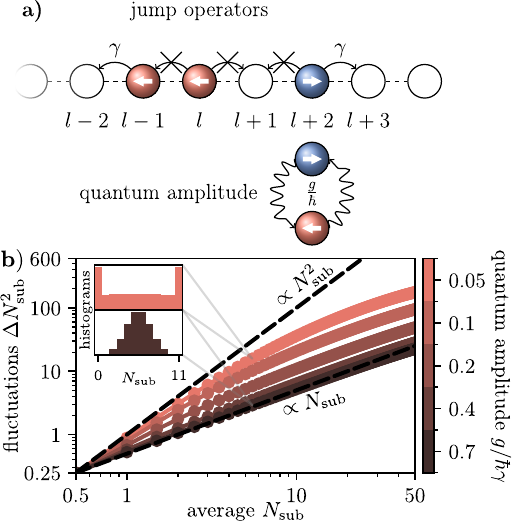}
    \caption{\textbf{a)} Schematic of dynamic processes in the minimal model.
    Particles jump only in the direction of their internal spin (no left jumps for $\rightarrow$ particles, and vice versa) and only if the target site is empty.
    The jump from $l$ to $l-1$ is blocked since $l-1$ is occupied.
    \textbf{b)} Giant number fluctuations for different subsystem sizes.
    Total system size $\systemsize=1000$, $\numbertrajectories={5\cdot 10^4}$ independent trajectories, one spatial configuration sampled per trajectory.
    Insets show histograms of $N_\mathrm{sub}$ for $L_\mathrm{sub} = 11$.
    The exponent for $g/\hbar\gamma = 0.05$ is $\beta \approx 1.7$.
    }
 \label{fig1:Model}
\end{figure}

\paragraph{Active Quantum Matter Model.}
\label{sec:model}

We consider a one-dimensional chain of $\systemsize$ sites with periodic boundary conditions, occupied by hard-core bosons with an internal spin degree of freedom $\sigma\in\{\leftarrow,\rightarrow\}$, which amounts to bosons with infinite on-site repulsion. We further impose this infinite repulsion across spin species, so that the local Hilbert space allows at most one particle per site regardless of its internal state. The system's density matrix $\rho$ evolves according to a Lindblad master equation
\begin{equation} \label{eq:LindbladMasterEquation}
 \frac{d\rho}{dt} = -\frac{i}{\hbar}[\ham, \rho] + \gamma \sum_{l,\sigma} \left( M_{l\sigma} \rho M_{l\sigma}^\dagger - \frac{1}{2}\{M_{l\sigma}^\dagger M_{l\sigma}, \rho\} \right),
\end{equation}
with the Hamiltonian $H$ and quantum jump operators $M_{l\sigma}$.
The dynamics therefore consists of mainly two competing processes:

1. Directed hopping (run): The internal spin is coupled to the direction of motion through asymmetric jump operators that move $\leftarrow$ and $\rightarrow$ particles in opposite directions at rate $\gamma$:
\begin{equation} \label{eq:jump_operators}
\begin{split}
 M_{l\leftarrow} &= P_{l-1}^{\rightarrow}\bdleft{l-1}\bleft{l}~, \quad
 M_{l\rightarrow}= P_{l+1}^\leftarrow\bdright{l+1}\bright{l}~.
\end{split}
\end{equation}
The exclusion factors $P_l^\sigma=(1-\nls{l}{\sigma})$ enforce the aforementioned hard-core constraint, allowing hops only into empty sites. This dissipative dynamics breaks local detailed balance~\cite{khasseh2024, Sieberer2016} making the system active.

2. Coherent reorientation (tumble): The Hamiltonian
\begin{equation} \label{eq:non-interacting_Hamiltonian}
 \ham = \ham_0 = -g\sum_{l} \left( \bdleft{l}\bright{l} + \bdright{l}\bleft{l} \right)~,
\end{equation}
coherently rotates the internal spin, inducing transitions between $\leftarrow$ and $\rightarrow$ particles. This serves as the quantum analog of the tumbling mechanism in classical run-and-tumble dynamics by randomizing the direction of self-propulsion~\cite{Tailleur2008, CatesTailleur2013, Cates2015}.
Both mechanisms are schematically depicted in Fig.~\ref{fig1:Model}a).
The competition between $g$ and $\gamma$ sets the persistence length, i.e., the mean number of directed hops before reversal. Importantly, the quantum flip probability $P_\mathrm{flip}(t)=\sin^2(gt/\hbar)\simeq(gt/\hbar)^2$ grows \emph{quadratically} at short times (Rabi oscillations), in contrast to the linear growth of classical tumbling. This quantum feature already gives a hint that quantum properties might actually enhance the effective persistence, strengthening the tendency towards clustering.

The system is initialized in the maximally mixed state at fixed particle number $N = \systemsize/2$. We then solve the Lindblad master equation numerically (see below) and evolve the system up to long times into the steady state, which remains out of equilibrium due to the persistent activity induced by the jump operators.
%


\paragraph{Monte Carlo Wave-Function Method and Exact Factorization.}
\label{sec:mcwf}

We simulate the dynamics using the Monte Carlo wave-function method (MCWF)~\cite{Dalibard1992,Dum1992,Molmer1993,Daley2014}, which unravels the Lindblad master equation into an ensemble of $\numbertrajectories$ stochastic pure-state trajectories $\{\ket{\psi_n(t)}\}_{n=1,\dots,\numbertrajectories}$. Each trajectory evolves under the effective non-Hermitian Hamiltonian $
\ham_\text{eff} = \ham - \frac{i\hbar\gamma}{2}\sum_{l\sigma} M_{l\sigma}^\dagger M_{l\sigma},$ interrupted by discrete quantum jumps associated with the dissipative channels $\{M_{l\sigma}\}$. Physical expectation values are obtained by averaging over trajectories.

A crucial property of our model enables numerically exact simulations up to system sizes of $L=1000$: the effective Hamiltonian $\ham_\text{eff}$ commutes with all local occupation operators $n_l = \sum_\sigma n_{l\sigma}$, so the spatial configuration $\{n_l\}$ remains fixed during deterministic evolution, and only quantum jumps redistribute particles. This reduces the Hilbert space from dimension $3^L$ to $2^N$, where $N$ is the total particle number.

For the minimal model, the dynamics factorize even further. Both the effective Hamiltonian and the quantum state decompose into independent single-site contributions,
\begin{align} \label{eq:effective_Hamiltonian_decomposition}
 \ham_\text{eff}&= \sum_l \ham^l_\text{eff}~,\quad
    |\psi_n(t)\rangle = \bigotimes_l |\psi_n^l(t)\rangle~,
\end{align}
where $|\psi_n^l(t)\rangle$ describes the spin of the particle at site $l$, or equals $|\emptyset\rangle$ if the site is unoccupied. The state is thus fully described by $2N$ complex numbers, enabling numerically exact simulations for $L$ up to $1000$.

We evolve the dynamics to times $T_\mathrm{max} \gg \gamma^{-1}, \hbar/g$ and verify convergence of all observables. The maximally mixed initial state is sampled from a uniform distribution over all occupation and spin configurations at fixed density $\rho=1/2$. Further implementation details of the MCWF method are provided in the End Matter.

\begin{figure*}[t]
 \centering
 \includegraphics[width=0.98\linewidth]{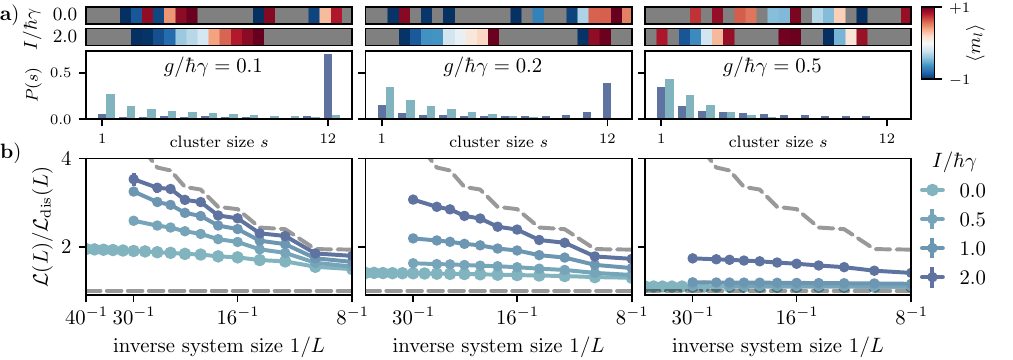}
 \caption{\textbf{a)} State of a representative trajectory of the steady-state for the minimal model (top) and the extended model with $I/\hbar\gamma=2$ (bottom), both for system size $\systemsize = 24$. The colors indicate the local spin magnetization $\langle m_l \rangle = \langle n_{l\leftarrow} - n_{l\rightarrow}\rangle$, where the expectation value is taken over the trajectory's spin degree of freedom. Below are histograms of cluster sizes $s$, defined as the number of consecutive occupied sites. The colors correspond to the interaction strength in \textbf{b)} for $I/\hbar\gamma = 0.0, 2.0$.
 At $g/\hbar\gamma=0.1$, the extended model exhibits a single large cluster, while the minimal model still shows multiple smaller clusters. For larger quantum amplitudes the cluster size distribution of both models become similar.
 \textbf{b)} Characteristic length scale $\mathcal{L}$ [Eq.~\eqref{eq:length_scale}] normalized by the disordered state length scale $\mathcal{L}_\mathrm{dis}(\systemsize)$ for different quantum amplitudes $g/\hbar\gamma$ (compare with \textbf{a)}) and interaction strengths $I/\hbar\gamma$.
 The dashed lines indicate the lower bound $\mathcal{L}_\mathrm{dis}(\systemsize)$ and an upper bound $\mathcal{L}_\mathrm{max}(\systemsize)$ obtained from a maximally clustered state.
 All points were acquired with $\numbertrajectories = 10^5$ independent trajectories.
 }
 \label{fig2:length_scale}
\end{figure*}

\paragraph{Giant Number Fluctuations.}
\label{sec:fluctuations}

Having established the exact factorization that enables numerically exact simulations up to $L=1000$, we now turn to the steady-state properties of the minimal model. A hallmark of MIPS is the presence of giant number fluctuations~\cite{Fily2012,Fily2014,Cates2015}, where the variance of the particle number in a subsystem grows superlinearly with its mean occupation.

For a connected subsystem $\mathcal{S}_\mathrm{sub}$ of size $L_\mathrm{sub}$, we define $N_\mathrm{sub} = \sum_{l\in \mathcal{S}_\mathrm{sub},\sigma} n_{l\sigma}$. In a homogeneous state with uncorrelated site occupations, $\Delta N_\mathrm{sub}^2 \propto \overline{N_\mathrm{sub}}$, while a fully phase-separated state gives $\Delta N_\mathrm{sub}^2 \propto \overline{N_\mathrm{sub}}^2$, since the probability distribution of the local density becomes bimodal and the variance is dominated by the density contrast between the coexisting dense and dilute phases~\cite{Cates2015,Fily2012}.

Exploiting the factorization~\eqref{eq:effective_Hamiltonian_decomposition}, we access subsystem sizes large enough to test this scaling directly up to $L = 1000$. As shown in Fig.~\ref{fig1:Model}b), for large quantum amplitudes $g/\hbar\gamma$, the number fluctuations are consistent with linear scaling, as expected in the weakly clustered regime. For weak quantum amplitudes, however, we observe a pronounced superlinear behavior $\Delta N_\mathrm{sub}^2 \propto \overline{N_\mathrm{sub}}^{\beta}$ with $\beta \approx 1.7$ for $g/\hbar\gamma = 0.05$, characteristic of strong clustering. As we will see, this clustering is accompanied by a finite characteristic length scale, motivating a microscopic analysis beyond number fluctuations alone.


\paragraph{Characteristic Length Scale.}
\label{sec:length_scale}

To test whether the superlinear fluctuations correspond to genuine phase separation, we analyze the characteristic length scale of density inhomogeneities~\cite{Gibaud2009,Henkes2011,Fily2012,Fily2014,Stenhammer2013,Smeets2016}. We define $\mathcal{L}$ from the static structure factor $S(q) = \sum_r e^{-iqr} C(r)$, where $C(r) = \frac{1}{\systemsize}\sum_l \langle n_l n_{l+r}\rangle$ is the density-density correlator, as
\begin{equation} \label{eq:length_scale}
 \mathcal{L} = 2\pi \frac{\int_{q_\mathrm{min}}^{q_\mathrm{max}} S(q)\,\mathrm{d}q}{\int_{q_\mathrm{min}}^{q_\mathrm{max}} S(q)\, q \,\mathrm{d}q}~,
\end{equation}
with $q_\mathrm{min} = 2\pi/\systemsize$ and $q_\mathrm{max}=\pi$. This quantity measures the typical size of density inhomogeneities: in a homogeneous state it approaches a finite value of order one lattice spacing, while for a phase-separated state it grows with system size.

For a disordered system, $C(r) = \rho_0^2 + \delta_{r0}(\rho_0 - \rho_0^2)$, yielding $\mathcal{L}_\mathrm{dis}(\systemsize) = 4/(1+2\systemsize^{-1})$, independent of the average density $\rho_0$.
Similarly, a maximally clustered system can be constructed by choosing all particles to be in a contiguous cluster and its length scale $\mathcal{L}_\mathrm{max}$ can be calculated numerically.

For the minimal model, the length scale $\mathcal{L}$ reveals two regimes, see Fig.~\ref{fig2:length_scale}. In the weakly clustered regime (large quantum amplitude $g/\hbar\gamma$), $\mathcal{L}$ converges to the homogeneous limit, consistent with the linear number fluctuations seen above. In the strongly clustered regime (weak quantum amplitude $g/\hbar\gamma$), $\mathcal{L}$ grows with system size. An extrapolation, however, is challenging because this growth appears to be very slow (slower than logarithmic), so that it might also finally converge to a finite value as $\systemsize\to\infty$, see Fig.~\ref{fig2:length_scale}. In the limit $g\to 0$ we do find evidence for a divergent length scale.

We therefore conclude that the minimal model exhibits pronounced clustering and superlinear number fluctuations, but realizes genuine phase separation only in the singular limit of vanishing quantum amplitude. The weak-$g$ regime is best described as \emph{microphase separation}~\cite{Leibler1980, Tjhung2018, Caporusso2020, Adachi2022}: the steady state organizes into dense clusters of finite characteristic size rather than a single macroscopic domain, as illustrated by the representative snapshot and cluster-size histogram in Fig.~\ref{fig2:length_scale}a). To construct these, we sample steady-state configurations and identify clusters as contiguous blocks of occupied sites. The cluster-size histogram then records the probability distribution of these block lengths. In the minimal model, the distribution remains concentrated at small sizes, consistent with finite clusters rather than coarsening into a single domain. This is reminiscent of the arrested coarsening and cluster phases reported for classical active particles~\cite{Buttinoni2013, Tjhung2018, Caporusso2020}.

\begin{figure}[t]
    \centering
    \includegraphics[width=\linewidth]{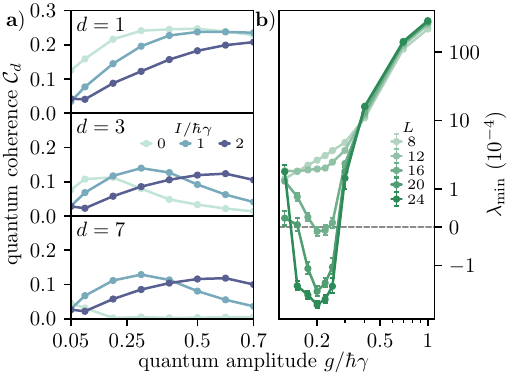}
    \caption{\textbf{a)} Quantum coherence $\mathcal{C}_d$ for $\systemsize=20$ as a function of quantum amplitude $g/\hbar\gamma$ for subsystem distances $d=1,3,7$ (vertical panels) and interaction strengths $I/\hbar\gamma = 0,1,2$ (curves, light to dark).
    For the minimal model, the coherence for weak quantum amplitudes remains finite even at large distances $d$, whereas it decays rapidly for strong quantum amplitudes.
    For the extended model, the coherence generally remains finite even at large distances for a broad range of quantum amplitudes, although slightly suppressed when compared to the minimal model.
    \textbf{b)} Smallest eigenvalue $\lambda_\mathrm{min}$ of the partial transpose of the nearest-neighbor reduced density matrix for $I/\hbar\gamma = 2$.
    Computed from $\numbertrajectories=10^4$ trajectories per data point.
    }
    \label{fig:quantum_coherence}
\end{figure}


\paragraph{Extended Model.}
\label{sec:interacting}

To explore whether robust quantum MIPS can be stabilized, we extend the Hamiltonian by a nearest-neighbor interaction $\ham = \ham_0 + \ham_\mathrm{I},
 \ham_\mathrm{I} = I \sum_{l\sigma} \nls{l}{\sigma} \nls{l+1}{\sigma}~.$
For $I>0$, this term implements a repulsive interaction between neighboring particles of the same spin.
It penalizes locally aligned motility states and favors spin anti-alignment on adjacent sites.
Although this favors spins pointing both toward and away from one another, motility breaks this symmetry and moves the particles apart when they are pointing away from each other.
The only long-lived configuration is therefore the one where the spins are pointing toward each other, thereby enhancing the self-trapping mechanism underlying MIPS.

This spin-dependent interaction is chosen to minimally break the exact factorization that made the minimal model solvable: a density-density interaction would preserve the factorization~\eqref{eq:effective_Hamiltonian_decomposition}, keeping the model tractable but not fundamentally altering its character. By breaking the factorization, the spin-dependent term introduces the genuine many-body coupling necessary to stabilize macroscopic phase separation, as we will provide evidence for.

The simulations become computationally more demanding without the factorization. While the reduction to fixed spatial configurations remains valid (since $\ham_\mathrm{eff}$ still commutes with $n_l$), the wave function no longer factorizes into independent local spin states. This restricts our numerically exact MCWF simulations to system sizes up to $\systemsize\approx30$, which are not of sufficient size for reliable number-fluctuation scaling. We therefore focus on the characteristic length scale $\mathcal{L}$ as the central diagnostic.

Figure~\ref{fig2:length_scale}b) shows $\mathcal{L}$ for the extended model alongside the minimal model results. The difference is compelling: for weak quantum amplitudes, $\mathcal{L}$ in the extended model approaches the upper bound a maximally clustered state would satisfy, whereas the minimal model showed only a slow growth with system size, which could also be seen consistent with a finite asymptotic value. This is further supported by the representative snapshots and cluster-size histograms in Fig.~\ref{fig2:length_scale}a): at $g/\hbar\gamma=0.1$, the particles in the extended model condense into a single dense cluster, while the minimal model still exhibits a broad distribution of smaller clusters. For large quantum amplitudes $g$, both models converge to the homogeneous limit.

The strong growth of $\mathcal{L}$ with system size in the nearest-neighbor interaction regime is consistent with genuine phase separation. Together with the superlinear number fluctuations found in the minimal model, these numerical results support that repulsive nearest-neighbor interactions drive the system into a regime compatible with one-dimensional quantum MIPS.

A recent work has proposed that a quantum analogue of MIPS could also be realized theoretically in the ground state of a non-Hermitian Hamiltonian~\cite{Adachi2022}.
However, whether such ground states can be experimentally realized remains unclear, so that we consider the present evidence in a Lindblad master equation setting as the first realistic setting for an actual quantum MIPS phase.

\paragraph{Quantum Coherence.}
\label{sec:coherence}

Having identified evidence for quantum MIPS, we now turn to another central question: does the activity that drives clustering destroy quantum coherence, or can the two coexist and even cooperate? To probe this, we compute the $\ell_1$ coherence~\cite{Baumgratz2014,Winter2016,Streltsov2017},
\begin{equation} \label{eq:quantum_coherence}
    \mathcal{C}_d = \sum_{i\neq j} |(\rho_d)_{ij}|~,
\end{equation}
which quantifies the relevance of superpositions by summing the off-diagonal elements of the two-site reduced density matrix $\rho_d = \mathrm{Tr}_{\overline{\{0,d\}}}(\rho) = \frac{1}{\numbertrajectories}\sum_n \mathrm{Tr}_{\overline{\{0,d\}}}\left(|\psi_n\rangle\langle\psi_n|\right),$
evaluated in the local spin-occupation basis $\{|\emptyset\rangle, |{\leftarrow}\rangle, |{\rightarrow}\rangle\}^{\otimes 2}$. Here, $\mathrm{Tr}_{\overline{\{0,d\}}}$ denotes the partial trace over all sites except $0$ and $d$. The equality is exact in the limit $\numbertrajectories\to\infty$ by linearity of the partial trace, and translational invariance ensures $\rho_d$ depends only on the distance $d$ between the two sites.

We compute $\mathcal{C}_d$ for $\systemsize=20$ as a function of $g/\hbar\gamma$, $d$, and $I$, see Fig.~\ref{fig:quantum_coherence}a). In the minimal model, the nearest-neighbor coherence $\mathcal{C}_1$ is finite across all parameters, while longer-range coherence depends strongly on $g$: for weak quantum amplitudes it remains sizable out to $d\approx 7$, whereas for strong amplitudes it decays beyond $d=2$. Adding the repulsive interaction $I>0$ slightly suppresses $\mathcal{C}_1$ but makes the coherence more robust at large distances: $\mathcal{C}_d$ stays finite at large separations over a broad range of quantum amplitudes.

These results reveal that the dissipative dynamics driving clustering do not simply destroy quantum coherence. The microscopic mechanism is intuitive: in a dense cluster, most hops are blocked by the exclusion constraint, rendering the dissipative dynamics ineffective~\cite{Soto2014, Slowman2016}. This leaves room for the unitary spin dynamics to generate superpositions, even for very weak quantum amplitudes $g$, where clustering is strongest. The clustering itself thus creates a protected environment for coherence: particles that are densely packed due to MIPS remain in coherent superposition states across the cluster. In the strongly clustered regime, the range of coherence grows together with the phase-separation tendency, demonstrating that activity and coherence are not competing but cooperating to generate a genuinely new state of matter. The clustered steady states cannot be described as a classical phase-separated mixture—they are inherently quantum.

The coherence is accompanied by entanglement signatures that differ between the two models. In the minimal model, the factorized wave function precludes spatial entanglement between sites; the steady state is a classical mixture of product states. Nevertheless, the local spinors retain coherence that survives the trajectory average. In the extended model, the factorization breaks down and the many-body state can develop spatial entanglement. For both models, we find small but finite entanglement between the spin species, as measured by the negativity under spin partial transpose. The resulting data are provided in the End Matter.

For the extended model we furthermore find spatial entanglement at $I/\hbar\gamma = 2$.
Fig.~\ref{fig:quantum_coherence}b) shows the smallest eigenvalue of the partial transpose of a reduced density matrix $\rho_{d=1}$ on nearest-neighbors, where here the partial transpose is done over the different sites.
A negative eigenvalue, as can be observed between $g\in [0.15, 0.3]$, leads to a non-zero negativity indicative of spatial entanglement.

\paragraph{Conclusion and Outlook.}
We have found compelling numerical evidence for quantum motility-induced phase separation in a one-dimensional open quantum system. Crucially, the clustered steady states retain quantum coherence through a cooperative interplay between activity and dissipation establishing quantum MIPS as a genuinely nonequilibrium quantum state distinct from its classical counterpart.

Several directions follow naturally. On the theoretical side, the exact factorization of the minimal model generalizes straightforwardly to higher spatial dimensions while retaining its tractable product-state structure, making it a scalable platform for exploring quantum MIPS and potentially other active quantum matter phenomena — such as flocking~\cite{Vicsek1995, Toner1995} or nematic order~\cite{Narayan2007, Marchetti2013} — at unprecedented scales. Developing analytical approaches, such as a hydrodynamic theory~\cite{Marchetti2013, Toner2005, CatesTailleur2013, Sieberer2016} where, however, one would need to incorporate properly the cooperative interplay between quantum coherence and dissipation, would provide further insight into the character of the nonequilibrium steady state.

On the experimental side, the key ingredients of our model — engineered directed dissipation, coherent spin rotation, and hard-core constraints — are within reach, in principle, of ultracold quantum and Rydberg atomic systems~\cite{khasseh2024, Diehl2008, Verstraete2009}, offering a concrete route to observing quantum MIPS. More broadly, our work establishes active quantum matter as a platform for nonequilibrium quantum phases in which quantum coherence is not a passive bystander but a cooperating participant.


\begin{acknowledgments}
\paragraph{Data availability - } The data contained in all figures of this article is available on Zenodo~\cite{Zenodo}.

\paragraph{Acknowledgements - }
This work was supported by the research funding program ``Forschungspotenziale besser nutzen!'' of the University of Augsburg.
The authors gratefully acknowledge the resources on the ALCC HPC cluster of the Institute of Physics at the University of Augsburg and on the LiCCA HPC cluster of the University of Augsburg, co-funded by the Deutsche Forschungsgemeinschaft (DFG, German Research Foundation) -- Project-ID 499211671.
R.A. acknowledges funding from the European Union through the ERC Starting Grant ``Living Fluctuations'' (no. 101114584).
\end{acknowledgments}

\bibliography{sources}

\newpage
\appendix
\section{End Matter}

\paragraph{Hard-core bosons.}\label{sec:HardCoreBosons}

Hard-core bosons combine properties of both fermionic and bosonic systems, exhibiting fermion-like anticommutation relations that enforce an exclusion principle while maintaining bosonic commutation behavior across different sites or spin states.

The creation and annihilation operators $\bdls{l}{\sigma}, \bls{l}{\sigma}$ satisfy fermionic anticommutation relations on the same site $l$ with the same spin degree of freedom $\sigma$:
\begin{align}
\{\bls{l}{\sigma}, \bdls{l}{\sigma}\} &= 1\\
\{\bls{l}{\sigma}, \bls{l}{\sigma}\} &= \{\bdls{l}{\sigma}, \bdls{l}{\sigma}\} = 0
\end{align}
For different sites or different spin degrees of freedom, the operators obey boson-like commutation relations:
\begin{align}
[\bls{l}{\sigma}, \bdls{l'}{\sigma'}] &= \delta_{ll'}\delta_{\sigma\sigma'}(1 - 2n_{l\sigma}), \\
[\bls{l}{\sigma}, \bls{l'}{\sigma'}] &= [\bdls{l}{\sigma}, \bdls{l'}{\sigma'}] = 0 \quad \text{for  } l\neq l' \text{ or } \sigma\neq\sigma'
\end{align}
Here, $n_{l\leftarrow} = \bdleft{l}\bleft{l}$ and $n_{l\rightarrow} = \bdright{l}\bright{l}$ are the number operators for spin-left and spin-right particles at site $l$, respectively.
The hard-core constraint is automatically enforced through these commutation relations, ensuring that each site can host at most one particle of each spin type.
The further constraint of at most one particle per site is implemented by prohibiting the state $|{\lrarrow}\rangle$ and the corresponding creation operators $\bdleft{l}\bdright{l}$ and $\bdright{l}\bdleft{l}$.

\paragraph{Monte Carlo wave-function method.}
\label{sec:MonteCarloWaveFunctionMethod}

In the MCWF approach, the density matrix evolution is unraveled into stochastic pure-state trajectories~\cite{Dalibard1992,Dum1992,Daley2014}.
Between jumps, each trajectory evolves under the non-Hermitian effective Hamiltonian $H_\mathrm{eff}$, so that the state norm decays and thereby encodes the cumulative jump probability.
In our implementation, the minimal model is represented by the particle positions together with a spin wave function for the occupied sites.
Because $[H_\mathrm{eff},n_l]=0$ for $H_0$, the no-jump part does not change the spatial configuration; for a fixed configuration, the deterministic evolution reduces to independent local $2\times 2$ problems, whose eigendecompositions are precomputed and used for efficient propagation.
At the beginning of each jump cycle, a random number $r\in(0,1)$ is drawn.
The state is then propagated forward until a jump time $t_\mathrm{jump}$ is reached, at which point its norm satisfies $\langle \psi(t_\mathrm{jump})|\psi(t_\mathrm{jump})\rangle=r$.
The jump time is numerically located by the bisection method.
A jump channel $M_{l\sigma}$ is then chosen with probability
\begin{equation}
    P(M_{l\sigma}) = \frac{\langle \psi(t_\mathrm{jump})|M_{l\sigma}^\dagger M_{l\sigma}|\psi(t_\mathrm{jump})\rangle}{\sum_{l'\sigma'}\langle \psi(t_\mathrm{jump})|M_{l'\sigma'}^\dagger M_{l'\sigma'}|\psi(t_\mathrm{jump})\rangle}~,
\end{equation}
where, for efficiency reasons, the jump channels prohibited by occupation constraints can be excluded.
After the jump, the particle position is updated, the corresponding spinor is projected onto the post-jump basis state, the entire state is normalized, and the deterministic evolution resumes.
Observables are measured from the normalized state at the prescribed times and averaged over many independent trajectories and randomly sampled initial states.
While \cite{Suess2014} presents an alternative algorithm, we found this simpler method to be more efficient for our system.

\paragraph{Convergence for quantum coherence.}

\begin{figure}
    \centering
    \includegraphics[width=\linewidth]{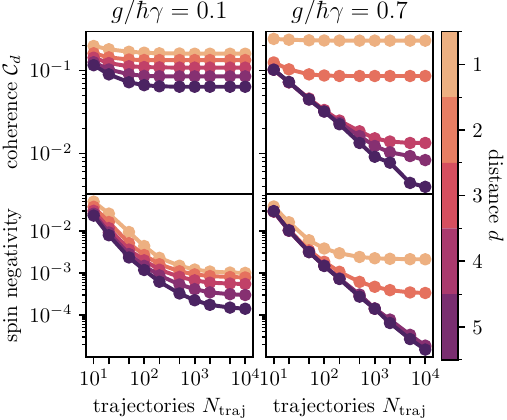}
    \caption{
    Top row: Convergence of quantum coherence \eqref{eq:quantum_coherence} of the reduced density matrix $\rho_d$ for $L=20$ at $I=0$.
    Bottom row: Convergence of spin negativity.
    For larger quantum amplitudes $g/\hbar\gamma$ and distances $d$ convergence cannot yet be observed for the $\numbertrajectories = 10^4$ trajectories.
    In this case, we use the value at $\numbertrajectories = 10^4$ as an upper bound estimate for the quantum coherence.
    }
    \label{fig:quantum_coherence_fit}
\end{figure}

We checked the convergence of the quantum coherence \eqref{eq:quantum_coherence} with respect to the number of trajectories $\numbertrajectories$.
Figure~\ref{fig:quantum_coherence_fit} shows the quantum coherence for $L=20$ at $I=0$ for different values of $g/\hbar\gamma$ and $d$.

\paragraph{Spin partial transpose and negativity.}

In the minimal model the many-body state is, along each trajectory, a product over sites [Eq.~\eqref{eq:effective_Hamiltonian_decomposition}].
The two-site reduced density matrix $\rho_d$ is therefore a classical mixture of product states and hence separable, so no spatial entanglement between sites can arise by construction.
Nevertheless, the local spinors retain coherence that survives the trajectory average, giving $\mathcal{C}_d>0$: the steady state is spatially unentangled yet genuinely quantum, which identifies coherence---rather than entanglement---as the relevant quantum signature.
In the extended model the factorization breaks down and spatial entanglement between sites can occur.

Since spatial entanglement in the reduced two-site density matrices is absent in the minimal model, we focus on entanglement in the spin degree of freedom.
To probe this, we apply a partial transpose in spin space to the reduced density matrix $\rho_d$, that we represent in terms of the spin occupation numbers on the two-site subsystem as
\begin{equation}
    \rho_d = \sum_{\mathbf{n},\mathbf{n}'}(\rho_d)_{\mathbf{n}, \mathbf{n}'}|n_{0\leftarrow}n_{0\rightarrow}n_{d\leftarrow}n_{d\rightarrow} \rangle\langle n'_{0\leftarrow}n'_{0\rightarrow}n'_{d\leftarrow}n'_{d\rightarrow}|~,
\end{equation}
where $\mathbf{n} = (n_{0\leftarrow}, n_{0\rightarrow}, n_{d\leftarrow}, n_{d\rightarrow})$ denotes the occupation numbers of the two spin states at sites $0$ and $d$.
For this consideration, we also need to include the doubly occupied state $|{\lrarrow}\rangle \coloneqq |1_{\leftarrow}1_{\rightarrow}\rangle$, since the partial transpose can mix states with different particle numbers.
The partial transpose in spin space acts by exchanging the $\rightarrow$-spin indices
\begin{equation}
    \rho_d^{T_\rightarrow} = \sum_{\mathbf{n},\mathbf{n}'}(\rho_d)_{\mathbf{n}, \mathbf{n}'}|n_{0\leftarrow}n'_{0\rightarrow}n_{d\leftarrow}n'_{d\rightarrow} \rangle\langle n'_{0\leftarrow}n_{0\rightarrow}n'_{d\leftarrow}n_{d\rightarrow}|~.
\end{equation}
This can be pictured by placing the $\leftarrow$ and $\rightarrow$ particles on the upper and lower legs of a two-rung ladder, respectively, and the partial transpose transposes the lower leg only.
According to the Peres--Horodecki criterion~\cite{Peres1996,Horodecki1996}, negative eigenvalues of the partially transposed matrix constitute a sufficient condition for entanglement between the two legs, i.e.\ between the two spin species.
We thus calculate the negativity $\mathcal{N}_d$ of the partially transposed matrix $\rho_d^{T_\rightarrow}$, which is defined as the sum of the absolute values of its negative eigenvalues.
The results can be seen in the bottom row of Fig.~\ref{fig:quantum_coherence_fit}.
Even though the negativity is very small, it is clearly converged in the number of trajectories, and we can conclude that the spin degree of freedom is indeed entangled in the steady state.

\end{document}